\documentclass[conference,letter]{IEEEtran}
\IEEEoverridecommandlockouts

\usepackage{color}

\usepackage{graphicx,tabularx,array,amsmath,amsthm,thmtools}

\usepackage{mathtools}

\usepackage{amsfonts}
\usepackage{bm}
\usepackage{bbm}
\usepackage{makecell}
\usepackage{multirow}
 \usepackage{amssymb}
\usepackage{txfonts}
\usepackage[T1]{fontenc}
\usepackage{tikz}
\usepackage[scr=dutchcal]{mathalfa}

\usepackage{cite}
\newcommand{\indep}{\rotatebox[origin=c]{90}{$\models$}}
\newtheorem{Theorem}{Theorem}
\newtheorem{Proposition}{Proposition}
\newtheorem{Lemma}{Lemma}
\newtheorem{Corollary}{Corollary}

\newtheorem{Remark}{Remark}
\newtheorem{Definition}{Definition}

\begin{document}
%
\title{Matching Graphs with Community Structure: A Concentration of Measure Approach}

\author{\IEEEauthorblockN{ Farhad Shirani}
\IEEEauthorblockA{Department of Electrical\\and Computer Engineering\\
New York University\\
New York, New York, 11201\\
Email: fsc265@nyu.edu}
\and

\IEEEauthorblockN{Siddharth Garg}
\IEEEauthorblockA{Department of Electrical\\and Computer Engineering\\
New York University\\
New York, New York, 11201\\ 
Email: siddharth.garg@nyu.edu}

\and

\IEEEauthorblockN{Elza Erkip }
\IEEEauthorblockA{Department of Electrical\\and Computer Engineering\\
New York University\\
New York, New York, 11201\\
Email: elza@nyu.edu}
}


%


\maketitle

\begin{abstract}
 In this paper, matching pairs of random graphs under the community structure model is considered. The problem emerges naturally in various applications such as privacy, image processing and DNA sequencing. A pair of randomly generated labeled graphs with pairwise correlated edges are considered. It is assumed that the graph edges are generated based on the community structure model. Given the labeling of the edges of the first graph, the objective is to recover the labels in the second graph. The problem is considered under two scenarios: i) with side-information where the community membership of the nodes in both graphs are known, 
 and ii) without side-information where the community memberships are not known.  A matching scheme is proposed which operates based on typicality of the adjacency matrices of the graphs. Achievability results are derived which provide theoretical guarantees for successful matching under specific assumptions on graph parameters. It is observed that for the proposed matching scheme, the conditions for successful matching do not change in the presence of side-information. Furthermore, a converse result is derived which characterizes a set of graph parameters for which matching is not possible. 
 \let\thefootnote\relax\footnotetext{This research was supported in part by NSF grants CNS-1553419 and CCF-1815821.}

\end{abstract}
\section{Introduction}
The graph matching problem emerges naturally in a wide range of applications including social network de-anonymization, pattern recognition, DNA sequencing, and database alignment. 
In this problem, an agent is given a correlated pair of randomly generated graphs: i) an `anonymized' unlabeled graph, and ii) a `de-anonymized' labeled graph. The objective is to leverage the correlation among the edges of the graphs to find the canonical labeling of the vertices in the anonymized graph. 

There has been extensive research investigating the fundamental limits of graph matching, i.e. characterizing the necessary and sufficient conditions for successful matching.  The problem has been considered under various probabilistic models capturing the correlation among the graph edges. In its simplest form - where the edges of the two graphs are exactly equal and are generated independently- it is called \textit{graph isomorphism} and has been studied in \cite{iso1,iso2,iso4}. The Erd\H{o}s-R\'enyi model provides a generalization where the edges in the two graphs are pairwise correlated and are generated independently, based on identical distributions. More precisely, in this model, edges whose vertices are labeled identically, are correlated through an arbitrary joint probability distribution and are generated independently of all other edges.
 Matching under the Erd\H{o}s-R\'enyi model was considered in  ~\cite{corr1,corr2,corr3,corr4, kia_2017, Lyzinski_2016, Asilomar, cullina2017exact, arxiv_matching_ISIT18}. The Erd\H{o}s-R\'enyi model allows for arbitrary but identical correlations among edge pairs in the two graphs. Consequently, it does not model the community structure among the graph nodes which manifests in many applications \cite{community}. As an example, in social networks, users may be divided into communities based on various factors such as age-group, profession, and racial background. The users' community memberships affects the probability that they are connected with each other. A matching algorithm may use the community membership information to enhance its performance. In order to take the users' community memberships into account, an extension to the Erd\H{o}s-R\'enyi model is considered which is called the \textit{community structure model}. In this model, the edge probabilities depend on their corresponding vertices' community memberships. There has several works studying graph matching schemes under the community structure model \cite{nilizadeh2014community,singhal2017significance}. 
 
 In this work, we consider the graph matching problem under the community structure model. We build upon the typicality matching scheme which was proposed in our prior work \cite{arxiv_matching_ISIT18} to construct a matching scheme under two scenarios: i) with side-information, where the community membership of the nodes in both graphs are given, 
 and ii) without side-information, where the community memberships are not known in either graph. We derive 
 necessary conditions on graph parameters under which successful matching is possible.  Furthermore, we derive a converse result which characterizes a set of graph parameters for which matching is not possible.

The rest of the paper is organized as follows: Section \ref{sec:prelim} provides the mathematcial tools and background used in the rest of the paper. Section \ref{sec:typ} includes a result on the joint  typicality of permutations of pairs of correlated sequences. Section \ref{sec:main} provides achievability results for graph matching under the community structure model. Section \ref{sec:converse} includes a converse matching result. Section \ref{sec:conclusion} concludes the paper. 



 



\section{Preliminaries}
\label{sec:prelim}
This section describes the graph matching problem and introduces the mathematical machinery used in the rest of the paper. The first part of the section provides a formal description of the graph matching problem under the community structure model. The second part provides the necessary background on the joint typicality of permutations of pairs of correlated sequences which is the basis for our proposed matching scheme.

\subsection{Problem Formulation}

We consider graphs whose edges take multiple values.  An edge which has an attribute assignment is called a \emph{marked edge}. The following defines an unlabeled graph with $c$ communities whose edges may take $l$ different values, where $c\in \mathbb{N}$ and $l\geq 2$.
\begin{Definition}[\bf{Graph with Community Structure}]
 An $(n,c,(n_i)_{i\in [c]},l)$-unlabeled graph with community structure (UCS) $g$ is the triple $(\mathcal{V},\mathcal{C},\mathcal{E})$, where $n,l,c,n_1,n_2,\cdots,n_c\in \mathbb{N}$ and $l\geq 2$. The set $\mathcal{V}=\{v_{1},v_{2},\cdots,v_{n}\}$ is called the vertex set. The set $\mathcal{C}=\{\mathcal{C}_{1},\mathcal{C}_{2},\cdots,\mathcal{C}_{c}\}$ provides a partition for $\mathcal{V}$ and is called the set of communities. The $i$th community is written as $\mathcal{C}_{i}= \{v_{j_1}, v_{j_2},\cdots, v_{j_{n_i}}\}$.
 The set $\mathcal{E}\subset\{(x,v_{j_1},v_{j_2})|x\in [0,l-1], j_1\in [1,n], j_2\in[1,n]\}$ is called the (marked) edge set of the graph. For the edge $(x,v_{j_1},v_{j_2})$, the variable `$x$' represents the value assigned to the edge between vertices $v_{j_1}$ and $v_{j_2}$. The set $\mathcal{E}_{i_1,i_2}=\{(x,v_{j_1},v_{j_2})\in \mathcal{E}|v_{j_1}\in \mathcal{C}_{i_1}, v_{j_2}\in \mathcal{C}_{i_2} \}$ is the set of edges connecting the vertices in communities $\mathcal{C}_{i_1}$ and $\mathcal{C}_{i_2}$. 
 
 \label{Def:unlabeled}
 
\end{Definition}

\begin{Remark}
In the context of Definition \ref{Def:unlabeled}, an unlabeled graph with binary valued edges is a graph for which $l=2$. In this case, if the pair $v_{n,i}$ and $v_{n,i}$ are not connected, we write $(0,v_{n,i},v_{n,j})\in \mathcal{E}$, otherwise $(1,v_{n,i},v_{n,j})\in \mathcal{E}$. 
\end{Remark}

\begin{Remark}
 Without loss of generality, we assume that for any arbitrary pair of vertices $(v_{n,i},v_{n,j})$, there exists a unique $x\in [0,l-1]$ such that $(x,v_{n,i},v_{n,j})\in \mathcal{E}$.
\end{Remark}

\begin{Remark}
In this work, we often consider sequences of graphs $g^{(n)}, n\in \mathbb{N}$, where $g^{(n)}$ has $n$ vertices. In such instances, we write  $g^{(n)}=(\mathcal{V}^{(n)},\mathcal{C}^{(n)},\mathcal{E}^{(n)})$ to characterize the $n$th graph in the sequence.
\end{Remark}

\begin{Definition}[\bf{Labeling}]
  For an $(n,c,(n_i)_{i\in [c]},l)$-UCS $g=(\mathcal{V},\mathcal{C},\mathcal{E})$, a labeling is defined as a bijective function $\sigma: \mathcal{V}\to [1,n]$.  
The pair $\tilde{g}=(g, \sigma)$ is called an $(n,c,(n_i)_{i\in [c]},l)$-labeled graph with community structure (LCS). For the labeled graph $\tilde{g}$ the adjacency matrix is defined as $\widetilde{G}_{\sigma}=[\widetilde{G}_{\sigma,i,j}]_{i,j\in [1,n]}$ where $\widetilde{G}_{\sigma,i,j}$ is the unique value such that $(\widetilde{G}_{\sigma,i,j},v_i,v_j)\in \mathcal{E}_n$, where $(v_i,v_j)=(\sigma^{-1}(i),\sigma^{-1}(j))$. The submatrix $\widetilde{G}_{\sigma,\mathcal{C}_i,\mathcal{C}_j}=[\widetilde{G}_{\sigma,i,j}]_{i,j: v_i,v_j\in \mathcal{C}_i\times \mathcal{C}_j}$
is the adjacency matrix corresponding to the community pair $\mathcal{C}_i$ and $\mathcal{C}_j$. The upper triangle (UT) corresponding to $\tilde{g}$ is the structure $\widetilde{G}^{UT}_{\sigma}=[\widetilde{G}_{\sigma,i,j}]_{i<j}$. The upper traingle corresponding to communities $\mathcal{C}_i$ and $\mathcal{C}_j$ in $\tilde{g}$ is denoted by $\widetilde{G}^{UT}_{\sigma,\mathcal{C}_i,\mathcal{C}_j}=[\widetilde{G}_{\sigma,i,j}]_{i<j: v_i,v_j\in \mathcal{C}_i\times \mathcal{C}_j}$. 
\end{Definition}

Any pair of labelings are related through a permutation as described below.
\begin{Definition}
For two labelings $\sigma$ and $\sigma'$, the $(\sigma,\sigma')$-permutation is defined as the bijection $\pi_{(\sigma,\sigma')}$, where:
\begin{align*}
\pi_{(\sigma,\sigma')}(i)=j, \quad \text{if} \quad {\sigma'}^{-1}(j)=\sigma^{-1}(i), \forall i,j\in [1,n].
\end{align*}
\label{def:above}
\end{Definition}
We consider graphs generated stochastically based on the community structure model. In this model, the probability of an edge between a pair of vertices is determined by their community memberships as described below.

\begin{Definition}[\bf{Random Graph with Community Structure}]
\label{Def:RCS}
Let $P_{X|C_i,C_o}$ be a conditional distribution defined on  $\mathcal{X}\times \mathcal{C}\times\mathcal{C}$, where $\mathcal{X}=[0,l-1]$ and $\mathcal{C}$ is defined in Definition \ref{Def:unlabeled}. A random graph with community structure (RCS) $g_{P_{X|C_i,C_o}}$ is a randomly generated  $(n,c,(n_i)_{i\in [c]},l)$-UCS with vertex set $\mathcal{V}$,  community set $\mathcal{C}$, and edge set $\mathcal{E}$, such that
\begin{align*}
 Pr((x,c_{j_1},c_{j_2})\in \mathcal{E})= P_{X|C_i,C_o}(x|\mathcal{C}_{j_1},\mathcal{C}_{j_2}), \forall x\in [0,l-1],
\end{align*}
where $c_{j_1},c_{j_2}\in \mathcal{C}_{j_1}\times \mathcal{C}_{j_2}$, and edges between different vertices are mutually independent. 

\end{Definition}

\begin{Remark}
 For a given pair of communities $(\mathcal{C}_{j_1}, \mathcal{C}_{j_2})$, the value of $P_{X|C_i,C_o}(x|\mathcal{C}_{j_1}, \mathcal{C}_{j_2})$ is the probability that a vertex in $\mathcal{C}_{j_1}$ is connected to the vertex in $\mathcal{C}_{j_2}$ by an edge taking value $x$. In this work, we only consider undirected graphs, as a result, $P_{X|C_i,C_o}(x|\mathcal{C}_{j_1}, \mathcal{C}_{j_2})=P_{X|C_i,C_o}(x| \mathcal{C}_{j_2}, \mathcal{C}_{j_1})$. The  results can be extended to directed graphs in a straightforward manner. 
\end{Remark}

\begin{Remark}
In Definition \ref{Def:RCS}, if $c=1$, then the random graph becomes an Erd\H{o}s-R\'{e}nyi graph.  
\end{Remark}

The objective in the graph matching problem is to match the vertices of a pair of correlated RCSs. Two edges in a pair of RCSs are correlated given that their corresponding vertices have the same labeling, the edges are independent otherwise. A pair of correlated RCSs is formally defined below.

\begin{Definition}[\bf{Correlated Pair of RCSs}]
Let $P_{X,X'|C_i,C_o,C'_i,C'_o}$ be a conditional distribution defined on $\mathcal{X}\times \mathcal{X}'\times\mathcal{C}\times\mathcal{C}\times\mathcal{C}'\times\mathcal{C}'$,
where $\mathcal{X}=\mathcal{X}'=[0,l-1]$ and $(\mathcal{C},\mathcal{C}')$ are a pair of community sets of size $c\in \mathbb{N}$. A correlated pair of random graphs with community structure (CRCS)  $\tilde{\underline{g}}_{P_{X,X'|C_i,C_o,C'_i,C'_o}}=(\tilde{g}_{P_{X|C_i,C_o}},\tilde{g}'_{P_{X'|C'_i,C'_o}})$ is characterized by: 
i) the pair of RCSs $(g_{P_{X|C_i,C_o}},g'_{P_{X'|C'_i,C'_o}})$, ii) the pair of labelings $(\sigma,\sigma')$ for the unlabeled graphs $(g_{P_{X|C_i,C_o}},g'_{P_{X'|C'_i,C'_o}})$, and iii) the probability distribution $P_{X,X'|C_i,C_o,C'_i,C'_o}$, such that: 
\\1)The graphs have the same set of vertices $\mathcal{V}=\mathcal{V}'$. 
\\2) For any two edges $e=(x,v_{j_1},v_{j_2}), e'=(x',v'_{j'_1},v'_{j'_2}),  x,x'\in [0,l-1]$, we have
 \begin{align*}
 &Pr\left(e\in \mathcal{E}, e'\in \mathcal{E}'\right)=
 \\&
\begin{cases}
P_{X,X'|C_i,C_o,C'_i,C'_o}(x,x'|\mathcal{C}_{j_1},\mathcal{C}_{j_2}, \mathcal{C}'_{j'_1},\mathcal{C}'_{j'_2})
,& \text{if } \sigma(v_{j_l})=\sigma'(v'_{j'_l})\\
P_{X|C_i,C_o}(x|\mathcal{C}_{j_1},\mathcal{C}_{j_2})P_{X'|C'_i,C'_o}(x|\mathcal{C}'_{j'_1},\mathcal{C}'_{j'_2}), & \text{Otherwise}
\end{cases},
\end{align*}
where $l\in \{1,2\}$, $v_{j_1},v_{j_2}\in \mathcal{C}_{j_1}\times \mathcal{C}_{j_2}$, and $v'_{j'_1},v'_{j'_2}\in \mathcal{C'}_{j'_1}\times \mathcal{C'}_{j'_2}$. 
\label{Def:CRCS}
\end{Definition}

\begin{Remark}
 In Definition \ref{Def:CRCS}, we have assumed that  both graphs have the same number of vertices. In other words, the vertex set for both graphs is  $\mathcal{V}= \mathcal{V}'= \{v_1,v_2,\cdots,v_n\}$. We further assume that the community memberships in both graphs are the same. In other words, we assume that $v_j\in \mathcal{C}_i \Rightarrow v'_{j'}\in \mathcal{C}'_i$ given that $\sigma(v_j)=\sigma'(v'_{j'})$ for any $j,j'\in [n]$ and $i\in [c]$. However, the results presented in this work can be extended to graphs with unequal but overlapping vertex sets and unequal community memberships.
\end{Remark}

In the graph matching problem, a pair of correlated random graphs are given, where the first graph is labeled and the second graph is not labeled. The objective is to identify the canonical labeling of the second graph based on the edge correlations. It is assumed that the matching algorithm has access to the edge statistics. Furthermore, it may or may not have access to the community memberships of the vertices in the two graphs. The following definitions formally describe the graph matching scenarios considered in this paper. 

\begin{Definition}[\bf{Graph Matching Problem}]
For a given sequence of conditional distributions $P^{(n)}_{X,X'|C_i,C_o,C'_i,C'_o}, n\in \mathbb{N}$, a graph matching problem is characterized by a pair of partially labeled graphs with community structure (PLCS) $\underline{g}_{P^{(n)}_{X,X'|C_i,C_o,C'_i,C'_o}}=(\tilde{g}_{P^{(n)}_{X|C_i,C_o}},{g}'_{P^{(n)}_{X'|C'_i,C'_o}})$  consisting of: i) the pair of unlabeled graphs with community structure $({g}_{P^{(n)}_{X|C_i,C_o}},{g}'_{P^{(n)}_{X'|C'_i,C'_o}})$, ii) a labeling $\sigma^{(n)}$ for the unlabeled graph $g_{P^{(n)}_{X|C_i,C_o}}$, such that there exists a labeling ${\sigma'}^{(n)}$ for the graph ${g'}_{P^{(n)}_{X'|C'_i,C'_o}}$ for which $(\tilde{g}_{P^{(n)}_{X|C_i,C_o}},{\tilde{g}'}_{P^{(n)}_{X'|C'_i,C'_o}})$ is a CRCS with joint distribution $P^{(n)}_{X,X'|C_i,C_o,C'_i,C'_o}$, where ${\tilde{g}'}_{P^{(n)}_{X'|C'_i,C'_o}}\triangleq ({g}'_{P^{(n)}_{X'|C'_i,C'_o}},{\sigma'}^{(n)})$.
\end{Definition}


\begin{Remark}
We assume that the size of the communities in the graph sequence grows linearly in the number of vertices. More precisely, let $\Lambda^{(n)}(i)\triangleq|\mathcal{C}^{(n)}_i|$ be the size of the $i$th community, we assume that $\Lambda^{(n)}(i)= \Theta(n)$ for all $i\in [c]$. Furthermore, we assume that the number of communities $c$ is constant in $n$.
\end{Remark}

\begin{Definition}[\bf{Matching Algorithm}]
\label{Def:match}
A matching algorithm is defined under the following two scenarios:
\begin{itemize}
\item{\textbf{With Side-information:} A matching algorithm operating with complete side-information is a sequence of functions $f^{CSI}_n: (\underline{g}_{P^{(n)}_{X,X'|C_i,C_o,C'_i,C'_o}},\mathcal{C}^{(n)},\mathcal{C}^{' (n)})\mapsto  \hat{\sigma}^{' (n)}, n\in \mathbb{N}$, where $\underline{g}_{P^{(n)}_{X,X'|C_i,C_o,C'_i,C'_o}}$ is a PLCS with $n$ vertices. 
}
\item{\textbf{Without Side-information:} A matching algorithm operating without side-information is a sequence of functions $f^{WSI}_n: \underline{g}_{P^{(n)}_{X,X'|C_i,C_o,C'_i,C'_o}}\mapsto  {\hat{\sigma}}^{' (n)}, n\in \mathbb{N}$.}
\end{itemize}
The output of a successful matching algorithm satisfies $P\left({\sigma'}^{(n)}(v'_{J^{(n)}})=\hat{\sigma'}^{(n)}(v'_{J^{(n)}})\right)\to 1$ as $n\to \infty$, where the random variable $J^{(n)}$ is uniformly distributed over $[1,n]$ and   ${\sigma'}^{(n)}$ is the labeling  for the graph $g'_{P^{(n)}_{X'|C'_i,C'_o}}$ for which $(\tilde{g}_{P^{(n)}_{X|C_i,C_o}},\tilde{g}'_{P^{(n)}_{X'|C'_i,C'_o}})$ is a CRCS, where $\tilde{g}'_{P^{(n)}_{X'|C'_i,C'_o}}\triangleq ({g}'_{P^{(n)}_{X'|C'_i,C'_o}},{\sigma'}^{(n)})$.
\label{def:algo}
\end{Definition}

\begin{Remark}
Note that the output of a  successful matching algorithm $\hat{\sigma'}^{(n)}$ does not necessarily satisfy $\hat{\sigma'}^{(n)}={\sigma'}^{(n)}$. In other words, the pair $(\tilde{g}_{P^{(n)}_{X|C_i,C_o}},\hat{g}'_{P^{(n)}_{X'|C'_i,C'_o}})$ is not necessarily a CRCS, where $\hat{g}'_{P^{(n)}_{X'|C'_i,C'_o}}\triangleq ({g}'_{P^{(n)}_{X'|C'_i,C'_o}},\hat{\sigma'}^{(n)})$. Rather, the algorithm finds the correct labeling for \textit{almost} all of the vertices in ${g}'_{P^{(n)}_{X'|C'_i,C'_o}}$. 
\end{Remark}

The following defines an achievable region for the graph matching problem.
\begin{Definition}[\bf{Achievable Region}]
 For the graph matching problem, a family of sets of distributions $\widetilde{P}=(\mathcal{P}_n)_{n\in \mathbb{N}}$ is said to be in the achievable region if for every sequence of distributions $P^{(n)}_{X,X'|C_i,C_o,C'_i,C'_o}\in \mathcal{P}_n, n\in \mathbb{N}$, there exists a matching algorithm. The maximal achievable family of sets of distributions is denoted by $\mathcal{P}^*$.  
 \end{Definition}

\subsection{Permutations and Typical Sequences}
We use standard results on the joint typicality of correlated sequences to propose schemes for matching pairs of correlated random graphs with community structure. In the following we provide a brief background on mathematical tools used in the rest of the paper. For a more detailed summary the reader is referred to \cite{arxiv_matching_ISIT18}. 

\begin{Definition}[\bf{Type}]
Let $\mathcal{X}=\{1,2,\cdots, |\mathcal{X}|\}$ be a given alphabet. The $|\mathcal{X}|$-length vector $\underline{T}(x^n)=(T_1(x^n),T_2(x^n), \cdots, T_{|\mathcal{X}|}(x^n))$ is called the type of the vector $x^n$ where $T_i(x^n)$ is the number of occurrences of the $i$th symbol in $x^n$, i.e. $T_i(x^n)=\sum_{j\in [n]}\mathbbm{1}(x_j=i), i\in [1,|\mathcal{X}|]$. Let $P_X$ be a probability distribution on $|\mathcal{X}|$. We write $\underline{T}(x^n)\stackrel{.}{=} n(P_X\pm \epsilon)$ if the following inequalities hold:
\begin{align*}
n(P_X(i)-\epsilon)\leq T_i(x^n)\leq n(P_X(i)+\epsilon), i\in [|\mathcal{X}|].     
\end{align*}
\end{Definition}

\begin{Definition}[\bf{Joint Type}]
For a pair of vectors $(x^n,y^n)$ defined on the alphabet $\mathcal{X}^n\times \mathcal{Y}^n$, the $|\mathcal{X}|\times|\mathcal{Y}|$ matrix $\underline{T}(x^n,y^n)$ is called the joint type of $(x^n,y^n)$, where $T_{i,j}(x^n,y^n), i,j \in [1,|\mathcal{X}|]\times [1,|\mathcal{Y}|]$ is the number of simultaneous occurrences of the $i$th symbol in $x^n$ and the $j$th symbol in $y^n$.
\end{Definition}

\begin{Definition}[\bf{Typicality}]
\label{Def:typ}
Let the pair of random variables $(X,Y)$ be defined on the probability space $(\mathcal{X}\times\mathcal{Y},P_{X,Y})$, where $\mathcal{X}$ and $\mathcal{Y}$ are finite alphabets. The $\epsilon$-typical set of sequences of length $n$ with respect to $P_{X,Y}$ is defined as:
\begin{align*}
&A_{\epsilon}^n(X,Y)=
\\&\Big\{(x^n,y^n): \Big|\frac{1}{n}N(\alpha,\beta|x^n,y^n)-P_{X,Y}(\alpha,\beta)\Big|\leq \epsilon, \forall (\alpha,\beta)\in \mathcal{X}\times\mathcal{Y}\Big\},\\
&=\Big\{(x^n,y^n): \underline{T}(x^n,y^n)\stackrel{.}{=} n(P_{X,Y}(\alpha,\beta)\pm \epsilon), \forall (\alpha,\beta)\in \mathcal{X}\times\mathcal{Y}\Big\}
\end{align*}
where $\epsilon>0$, $n\in \mathbb{N}$, and $N(\alpha,\beta|x^n,y^n)= \sum_{i=1}^n \mathbbm{1}\left((x_i,y_i)=(\alpha,\beta)\right)$.  
\end{Definition}  
\begin{Definition}[\bf{Permutation}]
\label{def:perm1}
A permutation on the set of numbers $[1,n]$ is a bijection $\pi:[1,n]\to [1,n]$. The set of all permutations on the set of numbers $[1,n]$ is denoted by $\mathcal{S}_n$. 
\end{Definition}
\begin{Definition}[\bf{Cycles}]
 A permutation $\pi \in \mathcal{S}_n, n\in \mathbb{N}$  is called a cycle if there exists $m\in [1,n]$ and $\alpha_1,\alpha_2,\cdots,\alpha_m\in [1,n]$ such that i) $\pi(\alpha_i)=\alpha_{i+1}, i\in [1,m-1]$, ii) $\pi(\alpha_n)=\alpha_1$, and iii) $\pi(\beta)=\beta$ if $\beta\neq \alpha_i, \forall i\in [1,m]$. The variable $m$ is called the length of the cycle. The element $\alpha$ is called a fixed point of the permutation if $\pi(\alpha)=\alpha$. We write $\pi=(\alpha_1,\alpha_2,\cdots,\alpha_m)$.
 The permutation $\pi$ is called a non-trivial cycle if $m\geq 2$. 
\end{Definition}

\begin{Lemma}\cite{isaacs}
 Every permutation $\pi \in \mathcal{S}_n, n\in \mathbb{N}$ has a unique representation as a product of disjoint non-trivial cycles.  
\end{Lemma}
\begin{Definition}
\label{def:perm2}
 For a given sequence $y^n\in \mathbb{R}^n$ and permutation $\pi\in \mathcal{S}_n$, the sequence $z^n=\pi(y^n)$ is defined as $z^n=(y_{\pi(i)})_{i\in [1,n]}$.
\end{Definition}
\begin{Definition}[\bf{Parameters of a Permutation and Standard Permutations}]
For a given $n,m,r\in \mathbb{N}$, and $1\leq i_1\leq i_2\leq \cdots\leq i_r\leq n$ such that $n=\sum_{j=1}^ri_j+m$, an $(m,r,i_1,i_2,\cdots,i_r)$-permutation is a permutation in $\mathcal{S}_n$ which has $m$ fixed points and $r$ disjoint cycles with lengths $i_1,i_2,\cdots,i_r$, respectively.

The $(m,r,i_1,i_2,\cdots,i_r)$-standard permutation is defined as the $(m,r,i_1,i_2,\cdots,i_r)$-permutation consisting of the cycles $(\sum_{j=1}^{k-1}i_j+1,\sum_{j=1}^{k-1}i_j+2,\cdots,\sum_{j=1}^{k}i_j), k\in [1,r]$. Alternatively, the $(m,r,i_1,i_2,\cdots,i_r)$-standard permutation is defined as:
\begin{align*}
\pi&=(1,2,\cdots,i_1)(i_1+1,i_1+2,\cdots,i_1+i_2)\cdots 
\\&(\sum_{j=1}^{r-1}i_j+1,\sum_{j=1}^{r-1}i_j+2,\cdots,\sum_{j=1}^{r}i_j)(n-m+1)(n-m+2)\cdots (n).
\end{align*}
\label{def:stan_perm}
\end{Definition}
The following proposition was proved in \cite{arxiv_matching_ISIT18}.
\begin{Proposition}
 Let $(X^n,Y^n)$ be a pair of i.i.d sequences defined on finite alphabets. We have:
\\ i) For an arbitrary permutation $\pi\in \mathcal{S}_n$, 
 \begin{align*}
 P((\pi(X^n),\pi(Y^n))\in A_{\epsilon}^n(X,Y))=P((X^n,Y^n)\in A_{\epsilon}^n(X,Y)).
\end{align*}
ii)  let $n,m,r,i_1,i_2,\cdots,i_r\in \mathbb{N}$ be permutation parameters as described in Definition \ref{def:stan_perm}. Let $\pi_1$ be an arbitrary $(m,r,i_1,i_2,\cdots,i_r)$-permutation  and let $\pi_2$ be the $(m,r,i_1,i_2,\cdots,i_r)$-standard permutation. Then, 
\begin{align*}
 P((X^n,\pi_1(Y^n))\in A_{\epsilon}^n(X,Y))=P((X^n,\pi_2(Y^n))\in A_{\epsilon}^n(X,Y)).
 \end{align*}
\end{Proposition}

 \section{Typicality of Permuted Sequences}
\label{sec:typ}
In this section, we study the typicality of permutations of pairs of correlated sequences. More precisely, let $(X^n,Y^n)$ be a pair correlated sequences of independent and identically distributed (i.i.d) random variables distributed according to $P_{X,Y}$ and let $\pi\in \mathcal{S}_n$ be an arbitrary permutation acting on $n$-length sequences. We provide bounds on the probability of joint typicality of the pair $(X^n,\pi(Y^n))$ with respect to the distribution $P_{X,Y}$. 

\begin{Theorem}
Let $(X^n,Y^n)$ be a pair of i.i.d sequences defined on finite alphabets $\mathcal{X}$ and $\mathcal{Y}$, respectively. For any permutation $\pi$ with $m\in [n]$ fixed points, the following holds:
 \begin{align}
  \label{eq:perm_bound}
  &P((X^n,\pi(Y^n))\in A_{\epsilon}^n(X,Y))  
  \\&\nonumber \qquad \qquad \leq 2^{-\frac{n}{4}(D(P_{X,Y}
 ||(1-\alpha)P_XP_Y+ \alpha P_{X,Y})-|\mathcal{X}||\mathcal{Y}|\epsilon+O(\frac{\log{n}}{n}))},
\end{align}
where $\alpha= \frac{m}{n}$, and $D(\cdot||\cdot)$ is the Kullback-Leibler divergence. 
\label{th:1}
\end{Theorem}

The proof is provided in the Appendix.
An alternative method for bounding the probability in Equation \eqref{eq:perm_bound} was presented in \cite{arxiv_matching_ISIT18}. The arguments provided in this paper lead to a significant simplification of the proof and can be extended to problems involving more than two sequences of random variables in a straightforward manner. 

\begin{Remark}
The upper bound in Equation \eqref{eq:perm_bound} goes to $0$ as $n\to \infty$ for any non-trivial permutation (i.e. $\alpha$ bounded away from one) and small enough $\epsilon$, as long as $X$ and $Y$ are not independent. 
\end{Remark}

\begin{Remark}
The exponent in Equation \eqref{eq:perm_bound} can be interpreted as follows: for the fixed points of the permutation ($\alpha$ fraction of indices), we have $Z_i=Y_i$. As a result, the joint distribution of the elements  $(X_i,Z_i)$ is $P_{X,Y}$. For the rest of the elements, $Z_i$ are permuted components of $Y^n$, as a result $(X_i,Z_i)$ are an independent pair of variables since $X^n$ and $Y^n$ are i.i.d. sequences. Consequently, the distribution of $(X_i,Z_i)$ is $P_XP_Y$ for $(1-\alpha)$ fraction of elements which are not fixed points of the permutation. The average distribution is $(1-\alpha)P_{X}P_{Y}+\alpha P_{X,Y}$ which appears in the exponent in Equation \eqref{eq:perm_bound}. 
\end{Remark}

\section{Matching Pairs of Correlated Graphs}
\label{sec:main}
In this section, we describe the typicality matching scheme and provide achievable regions for the the matching scenarios formulated in Definition \ref{Def:match}. 
\subsection{Matching in Presence of Side-information}
First, we describe the matching strategy under the complete side-information scenario.
In this scenario, the community membership of the nodes at both graphs are known prior to matching. Given a CRCS $\tilde{\underline{g}}_{P_{X,X'|C_i,C_o,C'_i,C'_o}}$, the scheme operates as follows. It finds a labeling ${\hat{\sigma}}'$, for which i) the set of pairs $(\widetilde{G}_{\sigma,\mathcal{C}_i,\mathcal{C}_j},\widetilde{G'}_{\hat{\sigma}',\mathcal{C}'_i,\mathcal{C}'_j}), i,j \in [c]$  are jointly typical each with respect to $P_{X,X'|C_i,C_o,C'_i,C'_o}(\cdot,\cdot|\mathcal{C}_i,\mathcal{C}_j,\mathcal{C}'_i,\mathcal{C}'_j)$ when viewed as vectors of length $n_i n_j, i\neq j$, and ii) the set of pairs $(\widetilde{G}^{UT}_{\sigma,\mathcal{C}_i,\mathcal{C}_i},\widetilde{G'}^{UT}_{\hat{\sigma}',\mathcal{C}'_i,\mathcal{C}'_i}), i \in [c]$ are jointly typical with respect to $P_{X,X'|C_i,C_o,C'_i,C'_o}(\cdot,\cdot|\mathcal{C}_i,\mathcal{C}_i,\mathcal{C}'_i,\mathcal{C}'_i)$ when viewed as vectors of length $\frac{n_i(n_i-1)}{2}, i\in [c]$. Specifically, it returns a randomly picked element $\hat{\sigma}'$ from the set:
\begin{align*}
 &\widehat{\Sigma}_{\mathcal{C}.\mathcal{C}'}=\{\hat{\sigma}'|
 (\widetilde{G}^{UT}_{\sigma,\mathcal{C}_i,\mathcal{C}_i}, \widetilde{G'}^{UT}_{\hat{\sigma}',\mathcal{C}'_i,\mathcal{C}'_i})\in A_{\epsilon}^{\frac{n_i(n_i-1)}{2}}(P_{X,X'|\mathcal{C}_i, \mathcal{C}_i,\mathcal{C}'_i, \mathcal{C}'_i}),\forall i\in [c],
 \\&
 (\widetilde{G}_{\sigma,\mathcal{C}_i,\mathcal{C}_j}, \widetilde{G'}_{\hat{\sigma}',\mathcal{C}'_i,\mathcal{C}'_j})\in A_{\epsilon}^{n_i  n_j}(P_{X,X'|\mathcal{C}_i, \mathcal{C}_j,\mathcal{C}'_i, \mathcal{C}'_j}),\forall i,j\in [c], i\neq j
 \},
\end{align*}
where $\epsilon=\omega(\frac{1}{n})$, and declares $\hat{\sigma}'$ as the correct labeling. We show that under this scheme,  the probability of incorrect labeling for any given vertex is arbitrarily small for large $n$.
\begin{Theorem}
\label{th:ach1}
For the typicality matching scheme, a given family of sets of distributions $\widetilde{P}=(\mathcal{P}^{(n)})_{n\in \mathbb{N}}$ is achievable, if for any constant $\delta>0$ and every sequence of distributions $P^{(n)}_{X,X'|C_i,C_o,C'_i,C'_o}\in \mathcal{P}_n,$ and community sizes $(n^{(n)}_1,n^{(n)}_2,\cdots, n^{(n)}_c), n\in \mathbb{N}$:
\begin{align}
&\nonumber \forall \alpha\in [0,1-\delta]: 4(1-\alpha)\frac{\log{n}}{n}\leq 
\max_{[\alpha_i]_{i\in [c]}\in \mathcal{A}_{\alpha}}
\\&\nonumber
 \sum_{i,j\in [c], i< j}
\frac{n^{(n)}_in^{(n)}_j}{n^2}\cdot
\\&\nonumber
D(P^{(n)}_{X,Y|\mathcal{C}_i,\mathcal{C}_j}
 ||(1-\beta_{i,j})P^{(n)}_{X|\mathcal{C}_i,\mathcal{C}_j}P^{(n)}_{Y|\mathcal{C}_i,\mathcal{C}_j}+ \beta_{i,j} P^{(n)}_{X,Y|\mathcal{C}_i,\mathcal{C}_j})
\\&\nonumber
+\sum_{i\in [c]}
 \frac{n^{(n)}_i(n^{(n)}_i-1)}{2n^2}\cdot\\&
 D(P^{(n)}_{X,Y|\mathcal{C}_i,\mathcal{C}_i}
 ||(1-\beta_i)P^{(n)}_{X|\mathcal{C}_i,\mathcal{C}_i}P^{(n)}_{Y|\mathcal{C}_i,\mathcal{C}_i}+ \beta_i P^{(n)}_{X,Y|\mathcal{C}_i,\mathcal{C}_i}),
\label{eq:th21}
\end{align}
as $n\to \infty$, where $\mathcal{A}_{\alpha}=  \{([\alpha_i]_{i\in [c]}): \alpha_i\leq \frac{n^{(n)}_i}{n}, \sum_{i\in [c]}\alpha_i=\alpha\}$, and $\beta_{i,j}= \frac{n^2}{n^{(n)}_in^{(n)}_j}\alpha_i\alpha_j, i,j\in [c]$ and $\beta_i=\frac{n\alpha_i(n\alpha_i-1)}{n^{(n)}_i(n^{(n)}_i-1)}, i\in [c]$. The maximal family of sets of distributions which are achievable using the typicality matching scheme with complete side-information is denoted by $\mathcal{P}_{full}$. 
\label{th:3}
\end{Theorem}
The proof is provided in the Appendix.

\begin{Remark}
Note that the community sizes $(n^{(n)}_1,n^{(n)}_2,\cdots, n^{(n)}_c), n\in \mathbb{N}$ are assumed to grow in $n$ such that $\lim_{n\to \infty}\frac{n_i^{n}}{n}>0$. 
\end{Remark}
Theorem \ref{th:3} leads to the following achievable region for matching of pairs of Erd\H{o}s-R\`enyi graphs (i.e. $c=1$).
\begin{Corollary}
For the typicality matching scheme, a given family of sets of distributions $\widetilde{P}=(\mathcal{P}^{(n)})_{n\in \mathbb{N}}$ is achievable, if for every sequence of distributions $P^{(n)}_{X,X'}\in \mathcal{P}_n, n\in \mathbb{N}$, and any constant $\delta>0$:
\begin{align*}
 \forall \alpha\in [0,1-\delta]:& 8(1-\alpha)\frac{\log{n}}{n}\leq 
\\& D(P^{(n)}_{X,Y}
 ||(1-\alpha)P^{(n)}_{X}P^{(n)}_{Y}+ \alpha P^{(n)}_{X,Y}),
\end{align*}
as $n\to \infty$. 
\end{Corollary}

\subsection{Matching in Absence of Side-information}
The scheme described in the previous section can be extended to matching graphs without community memberships side-information. In this scenario, it is assumed that the distribution $P_{X,X'|\mathcal{C}_i,\mathcal{C}_o,\mathcal{C}'_i,\mathcal{C}'_o}$ is known, but the community memberships of the vertices in the graphs are not known. In this case, the scheme sweeps over all possible possible community membership assignments of the vertices in the two graphs. For each community membership assignment, the scheme attempts to match the two graphs using the method proposed in the complete side-information scenario. If it finds a labeling which satisfies the joint typicality conditions, it declares the labeling as the correct labeling. Otherwise, the scheme proceeds to the next community membership assignment. More precisely, for a given community assignment $(\hat{\mathcal{C}},\hat{\mathcal{C}'})$, the scheme forms the following ambiguity set
\begin{align*}
 \widehat{\Sigma}_{\hat{\mathcal{C}},\hat{\mathcal{C}}'}&=\{\hat{\sigma}'|
 (\widetilde{G}^{UT}_{\sigma,
 \hat{\mathcal{C}}_i,\hat{\mathcal{C}}_i}, \widetilde{G'}^{UT}_{\hat{\sigma}',\hat{\mathcal{C}}'_i,\hat{\mathcal{C}}'_i})\in A_{\epsilon}^{\frac{n_i(n_i-1)}{2}}(P_{X,X'|\hat{\mathcal{C}}_i, \hat{\mathcal{C}}_i,\hat{\mathcal{C}}'_i, \hat{\mathcal{C}}'_i}),\forall i\in [c],
 \\&
 (\widetilde{G}_{\sigma,\hat{\mathcal{C}}_i,\hat{\mathcal{C}}_j}, \widetilde{G'}_{\hat{\sigma}',\hat{\mathcal{C}}'_i,\hat{\mathcal{C}'_j}})\in A_{\epsilon}^{n_i  n_j}(P_{X,X'|\hat{\mathcal{C}}_i, \hat{\mathcal{C}}_j,\hat{\mathcal{C}}'_i, \hat{\mathcal{C}}'_j}),\forall i,j\in [c], i\neq j
 \}.
\end{align*}
Define $\widehat{\Sigma}_{0}$ as follows:
\begin{align*}
  \widehat{\Sigma}_{0}= \cup_{(\hat{\mathcal{C}},
  \hat{\mathcal{C}')}\in \mathsf{C}  } \widehat{\Sigma}_{\hat{\mathcal{C}},\hat{\mathcal{C}'}}.
\end{align*}
where $\mathsf{C}$ is the set of all possible community membership assignments. The scheme outputs a randomly and uniformly chosen element of  $\widehat{\Sigma}_{0}$ as the correct labeling. The following theorem shows that the achievable region for this scheme is the same as the one described in Theorem \ref{th:ach1}.

\begin{Theorem}
Let $\mathcal{P}_0$ be the maximal family of sets of achievable distributions for the typicality matching scheme without side-information. Then, $\mathcal{P}_0= \mathcal{P}_{full}$.
\label{th:ach2}
\end{Theorem}
The proof is provided in the Appendix. 

\section{Converse Results}
\label{sec:converse}
In this section, we provide conditions on the graph parameters under which graph matching is not possible. Without loss of generality, we assume that $(\bf{\sigma},\bf{\sigma}')$ are a pair of random labelings chosen uniformly among the set of all possible labeling for the two graphs. The following theorem is proved in the appendix. 

\begin{Theorem}
\label{th:converse}
For the graph matching problem under the community structure model with complete side-information, the following provides necessary conditions  for successful matching:
\begin{align*}
n\log{n}&\leq \sum_{i,j \in [c], i<j}n_in_j I(X,X'|\mathcal{C}_i,\mathcal{C}_j, \mathcal{C}'_i\mathcal{C}'_j)
\\&+  \sum_{i \in [c]}\frac{n_i(n_i-1)}{2} I(X,X'|\mathcal{C}_i,\mathcal{C}_i, \mathcal{C}'_i,\mathcal{C}'_i), 
\end{align*}
where $I(X,X'|\mathcal{C}_i,\mathcal{C}_j, \mathcal{C}'_i\mathcal{C}'_j)$ is defined with respect to $P_{X,X'|\mathcal{C}_i,\mathcal{C}_j, \mathcal{C}'_i\mathcal{C}'_j}$.
\end{Theorem}
The proof is provided in the Appendix. For Erd\H{o}s-R\`enyi graphs, the following corollary is a direct consequence of Theorem \ref{th:converse}.
\begin{Corollary}
For the graph matching problem under the  Erd\H{o}s-R\`enyi model, the following provides necessary conditions  for successful matching:
\begin{align*}
\frac{2\log{n}}{n}&\leq   I(X,X').
\end{align*}
\end{Corollary}

 \section{Conclusion}
 \label{sec:conclusion}
 We have considered the problem of matching of correlated graphs under the community structure model. We have studied two matching scenarios: i) with side-information where the community membership of the nodes in both graphs are given, 
 and ii) without side-information where the community memberships are not known.
We have proposed a matching scheme which operates based on typicality of the adjacency matrices of the graphs. We have derived achievability results which provide theoretical guarantees for successful matching under specific assumptions on graph parameters. We have shown that the performance of the proposed scheme is the same with and without side-information.  Furthermore, we have provided a converse result which characterizes a set of graph parameters for which matching is not possible. 
\appendix

\subsection{Proof of Theorem \ref{th:1}}
Define the following partition for the set of indices $[1,n]$:

\begin{align*}
&\mathcal{A}_{0}= \{1, i_1+1, i_1+i_2+1, \cdots, \sum_{j=1}^{r-1} i_j +1\},\\
&\mathcal{A}_1= \{k| \text{k is even},~ \& ~k\notin \mathcal{A}_0, ~\&~ k\leq \sum_{i=1}^r i_j\},\\
&\mathcal{A}_2= \{k| \text{k is odd},~ \& ~k\notin \mathcal{A}_0, ~\&~ k\leq \sum_{i=1}^r i_j\},\\
&\mathcal{A}_3= \{k| k>\sum_{i=1}^r i_j\}.
\end{align*}
The set $\mathcal{A}_1$ is the set of indices at the start of each cycle in $\pi$, the sets $\mathcal{A}_2$ and $\mathcal{A}_3$ are the sets of odd and even indices which are not start of any cycles and $\mathcal{A}_4$ is the set of fixed points of $\pi$. Let $Z^n=\pi(Y^n)$. It is straightforward to verify that $(X_i,Z_i), i\in \mathcal{A}_j, j\in [3] $ are three sequences of independent and identically distributed variables which are distributed according to $P_{X}P_{Y}$.  The reason is that the standard permutation shifts elements of a sequence by at most one position, whereas the elements in the sequences $(X_i,Z_i), i\in \mathcal{A}_j, j\in [3] $ are at least two indices apart and are hence independent of each other (i.e. $Z_i\neq Y_i)$. Furthermore, $(X_i,Z_i), i\in \mathcal{A}_4$ is a sequence of independent and identically distributed variables which are distributed according to $P_{X,Y}$ since $Z_i=Y_i$. Let $\underline{T}_j, j\in [4]$ be the type of the sequence $(X_i,Z_i), i\in \mathcal{A}_j, j\in [4] $. We are interested in the probability of the event $(X^n,Z^n)\in \mathcal{A}_{\epsilon^n}(X,Y)$. From Definition \ref{Def:typ} this event can be rewritten as follows:
\begin{align*}
 &P((X^n,Z^n)\in \mathcal{A}_{\epsilon^n}(X,Y))
 \\& =P(\underline{T}(X^n,Y^n)\stackrel{.}{=} n(P_{X,Y}(\alpha,\beta)\pm \epsilon))
 \\&= P(\alpha_1\underline{T}_1+\alpha_2\underline{T}_2+\alpha_3\underline{T}_3+\alpha_4\underline{T}_4\stackrel{.}{=}n(P_{X,Y}(\alpha,\beta)\pm \epsilon)),
\end{align*}
where $\alpha_i= \frac{|\mathcal{A}_i|}{n}, i\in [4]$ and addition is defined element-wise. We have: 
\begin{align*}
 &P((X^n,Z^n)\in \mathcal{A}_{\epsilon^n}(X,Y))
 =\sum_{(\underline{t}_1,\underline{t}_2,\underline{t}_3,\underline{t}_4)\in \mathcal{T}} P(\underline{T}_i=\underline{t}_i, i\in [4]),
\end{align*}
where $\mathcal{T}= \{(\underline{t}_1,\underline{t}_2,\underline{t}_3,\underline{t}_4):\alpha_1\underline{t}_1+\alpha_2\underline{t}_2+\alpha_3\underline{t}_3+\alpha_4\underline{t}_4\stackrel{.}{=}n(P_{X,Y}(\alpha,\beta)\pm \epsilon)\}$. Using the property that for any set of events, the probability of the intersection is less than or equal to the geometric average of the individual probabilities, we have:
\begin{align*}
 &P((X^n,Z^n)\in \mathcal{A}_{\epsilon^n}(X,Y))
 \\&\leq \sum_{(\underline{t}_1,\underline{t}_2,\underline{t}_3,\underline{t}_4)\in \mathcal{T}} \sqrt[4]{\Pi_{i\in [4]}P(\underline{T}_i=\underline{t}_i)}.
\end{align*}
Since the elements  $(X_i,Z_i), i\in \mathcal{A}_j, j\in [4] $ are i.i.d, it follows from standard information theoretic arguments \cite{csiszarbook} that:

\begin{align*}
    & P(\underline{T}_i=\underline{t}_i) \leq 2^{-|\mathcal{A}_i|(D(\underline{t}_i||P_XP_Y)-|\mathcal{X}||\mathcal{Y}|\epsilon)}, i\in [3],
    \\&P(\underline{T}_4=\underline{t}_4) \leq 2^{-|\mathcal{A}_4|(D(\underline{t}_4||P_{X,Y})-|\mathcal{X}||\mathcal{Y}|\epsilon)}.
\end{align*}
We have, 
\begin{align*}
 &P((X^n,Z^n)\in \mathcal{A}_{\epsilon^n}(X,Y))
 \\&\leq \!\!\!\!\!\!\!\! \sum_{(\underline{t}_1,\underline{t}_2,\underline{t}_3,\underline{t}_4)\in \mathcal{T}}\!\!\!\!\!\!\!\! \sqrt[4]{2^{-n(\alpha_1D(\underline{t}_1||P_XP_Y)+\alpha_2D(\underline{t}_2||P_XP_Y)+\alpha_3D(\underline{t}_3||P_XP_Y)+\alpha_4D(\underline{t}_4||P_{X,Y})-|\mathcal{X}||\mathcal{Y}|\epsilon)}}\\
 &\stackrel{(a)}{\leq}
 \sum_{(\underline{t}_1,\underline{t}_2,\underline{t}_3,\underline{t}_4)\in \mathcal{T}} \sqrt[4]{2^{-n(D(\alpha_1\underline{t}_1+\alpha_2\underline{t}_2+\alpha_3\underline{t}_3+\alpha_4\underline{t}_4
 ||(\alpha_1+\alpha_2+\alpha_3)P_XP_Y+ \alpha_4P_{X,Y})-|\mathcal{X}||\mathcal{Y}|\epsilon)}}
 \\&
 =
 |\mathcal{T}| \sqrt[4]{2^{-n(D(P_{X,Y}
 ||(1-\alpha)P_XP_Y+ \alpha P_{X,Y})-|\mathcal{X}||\mathcal{Y}|\epsilon)}}
 \\&\stackrel{(b)}{\leq} 2^{-\frac{n}{4}(D(P_{X,Y}
 ||(1-\alpha)P_XP_Y+ \alpha P_{X,Y})-|\mathcal{X}||\mathcal{Y}|\epsilon+O(\frac{\log{n}}{n}))},
\end{align*}
where the (a) follows from the convexity of the divergence function and (b) follows by the fact that the number of joint types grows polynomially in $n$ \cite{csiszarbook}.

\subsection{Proof of Theorem \ref{th:ach1}}
Let $\epsilon_n= O(\frac{\log{n}}{n})$ be a sequence of positive numbers. Fix $n\in \mathbb{N}$ and let $\epsilon=\epsilon_n$. For a given labeling $\sigma''$, define the event $\mathcal{B}_{\sigma''}$ as the event that the sub-matrices corresponding to each community pair are jointly typical:
\begin{align*}
&\mathcal{B}_{\sigma''}:
 (\widetilde{G}^{UT}_{\sigma,\mathcal{C}_i,\mathcal{C}_i}, \widetilde{G'}^{UT}_{\sigma'',\mathcal{C}'_i,\mathcal{C}'_i})\in A_{\epsilon}^{\frac{n_i(n_i-1)}{2}}(P_{X,X'|{\mathcal{C}}_i, {\mathcal{C}}_i,{\mathcal{C}}'_i, {\mathcal{C}}'_i}),\forall i\in [c],
 \\&
 (\widetilde{G}_{\sigma,\mathcal{C}_i,\mathcal{C}_j}, \widetilde{G'}_{\sigma'',\mathcal{C}'_i,\mathcal{C}'_j})\in A_{\epsilon}^{n_i\cdot n_j}(P_{X,X'|{\mathcal{C}}_i, {\mathcal{C}}_j,{\mathcal{C}}'_i, {\mathcal{C}}'_j}),\forall i,j\in [c], i\neq j
 \},
\end{align*}
Particularly, $\beta_{\sigma'}$ is the event that the sub-matrices are jointly typical under the canonical labeling for the second graph. From standard typicality arguments it follows that:
\begin{align*}
P(\mathcal{B}_{\sigma'})\to 1 \quad \text{as}\quad n\to \infty.
\end{align*}
So, $P(\widehat{\Sigma}_{\mathcal{C}.\mathcal{C}'}=\phi)\to 0$ as $n\to \infty$ since the correct labeling is a member of the set $\widehat{\Sigma}_{\mathcal{C}.\mathcal{C}'}$.
Let $(\lambda_n)_{n\in \mathbb{N}}$ be an arbitrary sequence of numbers such that $\lambda_n= \Theta(n)$. We will show that the probability that a labeling in $\widehat{\Sigma}_{\mathcal{C}.\mathcal{C}'}$ labels $\lambda_n$ vertices incorrectly goes to $0$ as $n\to \infty$.
Define the following:
\begin{align*}
 \mathcal{E}=\{{\sigma'}^2\Big| ||\sigma^2-{\sigma'}^2||_1\geq \lambda_n\},
\end{align*}
where $||\cdot||_1$ is the $L_1$-norm. The set $\mathcal{E}$ is the set of all labelings which match more than $\lambda_n$ vertices incorrectly.

We show the following:
\begin{align*}
 P(\mathcal{E}\cap \widehat{\Sigma}_{\mathcal{C}.\mathcal{C}'}\neq \phi)\to 0, \qquad \text{as} \qquad n\to \infty.
 \end{align*}
We use the union bound on the set of all permutations along with Theorem \ref{th:1} as follows:
\begin{align*}
  &P(\mathcal{E}\cap \widehat{\Sigma}_{\mathcal{C}.\mathcal{C}'}\neq \phi)
  = P(\bigcup_{{\sigma''}: ||\sigma'-{\sigma''}||_1\geq \lambda_n}\{{\sigma''}\in  \widehat{\Sigma}_{\mathcal{C}.\mathcal{C}'}\})
  \\&\stackrel{(a)}{\leq} \sum_{k=\lambda_n}^{n}\sum_{{\sigma''}: ||\sigma'-{\sigma''}||_1=k} P(\sigma''\in  \widehat{\Sigma}_{\mathcal{C}.\mathcal{C}'})
  \\&\stackrel{(b)}{=} \sum_{k=\lambda_n}^{n}\sum_{{\sigma''}: ||\sigma'-{\sigma'}''||_1=k}
   P(\beta_{\sigma''})
\\&\stackrel{(c)}{\leq} \sum_{k=\lambda_n}^{n}\sum_{{\sigma'}^2: ||\sigma^2-{\sigma'}^2||_0=k} 2^{O(nlog{n}))} \times
\\&   \prod_{i,j\in [c], i< j}
   2^{-\frac{n_i\cdot n_j}{4}(D(P_{X,X'|\mathcal{C}_i,\mathcal{C}_j,\mathcal{C}'_i,\mathcal{C}'_j}
 ||(1-\beta_{i,j})P_{X|\mathcal{C}_i,\mathcal{C}_j}P_{X'|\mathcal{C}'_i,\mathcal{C}'_j}+ \beta_{i,j} P_{X,X'|\mathcal{C}_i,\mathcal{C}_j,\mathcal{C}'_i,\mathcal{C}'_j}))}\times 
 \\&  \prod_{i\in [c]}
 2^{-\frac{n_i(n_i-1)}{8}(D(P_{X,X'|\mathcal{C}_i,\mathcal{C}_i,\mathcal{C}'_i,\mathcal{C}'_i}
 ||(1-\beta_i)P_{X|\mathcal{C}_i,\mathcal{C}_i}P_{X'|\mathcal{C}'_i,\mathcal{C}'_i}+ \beta_i P_{X,X'|\mathcal{C}_i,\mathcal{C}_i,\mathcal{C}'_i,\mathcal{C}'_i})} 
   \\&\stackrel{(d)}{\leq}  \sum_{k=\lambda_n}^{n} {n \choose k}(!k)
 \max_{[\alpha_i]_{i\in [c]}\in \mathcal{A}}( 2^{-\frac{n^2}{4}(\Phi([\alpha_i]_{i\in [c]})
 +O(\frac{\log{n}}{n}))
 })
  \\&\leq 
  \max_{\alpha\in [0,1-\frac{\lambda_n}{n}]} 
  \max_{[\alpha_i]_{i\in [c]}}( 2^{-\frac{n^2}{4}(-(1-\alpha)\frac{\log{n}}{n}+\Phi([\alpha_i]_{i\in [c]})
 +O(\frac{\log{n}}{n}))
 }),
 \end{align*}
where $\mathcal{A}= \{([\alpha_i]_{i\in [c]}): \alpha_i\leq \frac{n_i}{n}, \sum_{i\in [c]}\alpha_i=\frac{n-\lambda_n}{n}\}$ and
\begin{align*}
 &   \Phi([\alpha_i]_{i\in [c]})=
 \sum_{i,j\in [c], i< j}n_in_j \cdot \\&
 D(P_{X,X'|\mathcal{C}_i,\mathcal{C}_j,\mathcal{C}'_i,\mathcal{C}'_j}
  || (1-\beta_{i,j}) P_{X|\mathcal{C}_i,\mathcal{C}_j} P_{X'|\mathcal{C}'_i,\mathcal{C}'_j}
 + \beta_{i,j} P_{X,X'|\mathcal{C}_i,\mathcal{C}_j,\mathcal{C}'_i,\mathcal{C}'_j})
  \\&+\sum_{i\in [c]}
\frac{n_i(n_i-1)}{2}\cdot
 \\&
 D(P_{X,X'|\mathcal{C}_i,\mathcal{C}_i,\mathcal{C}'_i,\mathcal{C}'_i}
 ||(1-\beta_i)P_{X|\mathcal{C}_i,\mathcal{C}_i}P_{X'|\mathcal{C}'_i,\mathcal{C}'_i}+ \beta_i P_{X,X'|\mathcal{C}_i,\mathcal{C}_i,\mathcal{C}'_i,\mathcal{C}'_i}),
\end{align*}
 and $\beta_{i,j}= \frac{n^2}{n_in_j}\alpha_i\alpha_j$ and $\beta_i= \frac{n\alpha_i(n\alpha_i-1)}{n_i(n_i-1)}$. Here, $\alpha_i$ is the number of fixed points in the $i$th community divided by $n$, and $\beta_i$ is the number of fixed points in $G^{' UT}_{\sigma'',\mathcal{C}'_i,\mathcal{C}'_i}$ divided by $\frac{n_i(n_i-1)}{2}$, and $\beta_{i,j}$ is the number of fixed points in $G^{' UT}_{\sigma'',\mathcal{C}'_i,\mathcal{C}'_j}$ divided by $n_i n_j$. Inequality (a) follows from the union bound, (b) follows from the definition of $ \widehat{\Sigma}_{\mathcal{C}.\mathcal{C}'}$, in (c) we have used Theorem \ref{th:1}, in (d) we have denoted the number of derangement of sequences of length $i$ by $!i$. Note that the right hand side in the (d) goes to 0 as $n\to \infty$ as long as \eqref{eq:th21} holds.

\subsection{Proof of Theorem \ref{th:ach2}}
The proof is similar to that of Theorem \ref{th:ach1}. We provide an outline. It is enough to show that $|\widehat{\Sigma}_{0}|$ has the same exponent as that of $|\widehat{\Sigma}_{\mathcal{C}.\mathcal{C}'}|$. To see this note that the size of the set of all community membership assignments $\mathsf{C}$ has an exponent which is $\Theta(n)$:

\begin{align*}
    |\mathsf{C}|\leq 2^{cn}.
\end{align*}
On the other hand, 
\begin{align*}
    |\widehat{\Sigma}_{0}|\leq |\mathsf{C}|\cdot |\widehat{\Sigma}_{\mathcal{C}.\mathcal{C}'}|\leq 2^{nc}\cdot 2^{\Theta(n\log{n})}=2^{\Theta(n\log{n})}.
\end{align*}
The rest of the proof follows by the same arguments as in Theorem \ref{th:ach1}.

\subsection{Proof of Theorem \ref{th:converse}}

For asymptotically large $n$, and $\epsilon>0$, let $G$ and $G'$ be the adjacency matrices of the two graphs under a pre-defined labeling. Let $\hat{\sigma}$ be the output of the matching algorithm. Let $\mathbbm{1}_{C}$ be the indicator of the event that the matching algorithm mislabels at most $\epsilon$ fraction of the vertices.  Note that $\hat{\sigma}$ is a function of $\sigma',G,G'$. So:

\begin{align*}
    &0=H(\hat{\sigma}|\sigma,G,G')
    \\&\stackrel{(a)}{=} 
    H(\sigma',\hat{\sigma},\mathbbm{1}_C|\sigma,G,G')- H(\sigma', \mathbbm{1}_C| \hat{\sigma},\sigma,G,G')
    \\&
    =  H(\sigma',\hat{\sigma},\mathbbm{1}_C|\sigma,G,G')-
    \\&
    H(\sigma' |\mathbbm{1}_C, \hat{\sigma},\sigma,G,G')
    - H( \mathbbm{1}_C| \hat{\sigma},\sigma,G,G')
    \\& 
    \stackrel{(b)}{\geq}
      H(\sigma',\hat{\sigma},\mathbbm{1}_C|\sigma,G,G')-
    H(\sigma' |\mathbbm{1}_C, \hat{\sigma},\sigma,G,G')-1
    \\&=
     H(\sigma',\hat{\sigma},\mathbbm{1}_C|\sigma,G,G')-
     \\&
    P(\mathbbm{1}_C=1) H(\sigma' |\mathbbm{1}_C=1, \hat{\sigma},\sigma,G,G')-
    \\&
    P(\mathbbm{1}_C=0) H(\sigma' |\mathbbm{1}_C=0, \hat{\sigma},\sigma,G,G')-1
    \\&\stackrel{(c)}{\geq} 
    H(\sigma',\hat{\sigma},\mathbbm{1}_C|\sigma,G,G')-
    \epsilon n\log{n}-
    P_e n\log{n}-1
    \\&\stackrel{(d)}{\geq}  H(\sigma'|\sigma,G,G')- (\epsilon+P_e)n\log{n}-1,
\end{align*}
where in (a) we have used the chain rule of entropy, in (b) we have used the fact that $\mathbbm{1}_C$ is binary, in (c) we define the probability of mismatching more than $\epsilon$ fraction of the vertices by $P_e$, and (d) follows from the fact that entropy is non-negative.
 As a result, $H(\sigma'|\sigma,G,G')\lesssim \epsilon n\log{n}$. We have,


\begin{align*}
    n\log{n}\approx \log{n!}= H(\mathbf{\sigma'})
    \approx I(\mathbf{\sigma}';\mathbf{\sigma}, G, G').
\end{align*}
We have:
\begin{align*}
n\log{n}& \approx
I(\mathbf{\sigma}';\mathbf{\sigma}, G, G')\\
& = I(\mathbf{\sigma}'; G')+
I(\mathbf{\sigma}';\mathbf{\sigma}, G |G')
\\&\stackrel{(a)}{=}
I(\mathbf{\sigma}';\mathbf{\sigma}, G |G')
\\&= I(\mathbf{\sigma}'; G|G')
+I(\mathbf{\sigma}'; G |G',\mathbf{\sigma})
\\&\stackrel{(b)}{=}
I(\mathbf{\sigma}'; G |G',\mathbf{\sigma})
\\& \stackrel{(c)}{\leq}
I(\mathbf{\sigma}',G'; G |\mathbf{\sigma})
\\& \stackrel{(d)}{=}
I(G'; G |\mathbf{\sigma}, \mathbf{\sigma}')
\\&\stackrel{(e)}{=} \sum_{i,j \in [c], i<j}n_in_j I(X,X'|\mathcal{C}_i,\mathcal{C}_j, \mathcal{C}'_i\mathcal{C}'_j)
\\&+  \sum_{i \in [c]}\frac{n_i(n_i-1)}{2} I(X,X'|\mathcal{C}_i,\mathcal{C}_i, \mathcal{C}'_i,\mathcal{C}'_i),
\end{align*}
where (a) follows from $\sigma'\indep G'$, (b) follows from the fact that $\sigma'\indep G,G'$, (c) is true due to the non-negativity of the mutual inforamtion, (d) follows from $\sigma,\sigma'\indep G$, and (e) follows from the fact that the edges whose vertices have different labels are independent of each other given the labels.

\IEEEpeerreviewmaketitle

%
%

\bibliographystyle{IEEEtran}

\end{document}